\documentclass[conference]{IEEEtran}
\IEEEoverridecommandlockouts
\usepackage{cite}
\usepackage{amsmath,amssymb,amsfonts}
\usepackage{algorithmic}
\usepackage{graphicx}
\usepackage{textcomp}
\usepackage{xcolor}
\usepackage{multirow}
\usepackage{makecell}
\usepackage{subfigure}
\usepackage{amsthm}
\usepackage{siunitx}
\usepackage[linesnumbered,ruled,vlined]{algorithm2e}
\usepackage{verbatim}
\def\BibTeX{{\rm B\kern-.05em{\sc i\kern-.025em b}\kern-.08em
    T\kern-.1667em\lower.7ex\hbox{E}\kern-.125emX}}
\makeatletter
\newcommand{\removelatexerror}{\let\@latex@error\@gobble}
\makeatother
\begin{document}

\title{\Large\textbf{FuseFPS: Accelerating Farthest Point Sampling with Fusing KD-tree Construction for Point Clouds}\\~\\
}	

\author{
Meng~Han, Liang~Wang, Limin~Xiao, Hao~Zhang, Chenhao~Zhang, Xilong~Xie, Shuai~Zheng, Jin~Dong
}

\author{Meng~Han\textsuperscript{1}, Liang~Wang\textsuperscript{1}, Limin~Xiao\textsuperscript{1}, Hao~Zhang\textsuperscript{1}, Chenhao~Zhang\textsuperscript{1},  Xilong~Xie\textsuperscript{1}, Shuai~Zheng\textsuperscript{1} Jin~Dong\textsuperscript{2}\\
\textsuperscript{1} State Key Laboratory of Software Development Environment \\
School of Computer Science and Engineering \\ 
Beihang University, Beijing 100191, China \\
\{hanm, lwang20, xiaolm, zh19373126, zch13021728086, xxl1399, zhengshuai\}@buaa.edu.cn \\
\textsuperscript{2} Beijing Academy of Blockchain and Edge Computing \\ 
dongjin@baec.org.cn} 

\maketitle

\begin{abstract}
Point cloud analytics has become a critical workload for embedded and mobile platforms across various applications. Farthest point sampling (FPS) is a fundamental and widely used kernel in point cloud processing. However, the heavy external memory access makes FPS a performance bottleneck for real-time point cloud processing. Although bucket-based farthest point sampling can significantly reduce unnecessary memory accesses during the point sampling stage, the KD-tree construction stage becomes the predominant contributor to execution time. In this paper, we present FuseFPS, an architecture and algorithm co-design for bucket-based farthest point sampling. We first propose a hardware-friendly sampling-driven KD-tree construction algorithm. The algorithm fuses the KD-tree construction stage into the point sampling stage, further reducing memory accesses. Then, we design an efficient accelerator for bucket-based point sampling. The accelerator can offload the entire bucket-based FPS kernel at a low hardware cost. Finally, we evaluate our approach on various point cloud datasets. The detailed experiments show that compared to the state-of-the-art accelerator QuickFPS, FuseFPS achieves about 4.3$\times$ and about 6.1$\times$ improvements on speed and power efficiency, respectively.
\end{abstract}

\section{Introduction}

A point cloud is a collection of points that directly represent 3D scenes or objects. Point cloud-based machine perception has become increasingly prevalent in embedded and mobile platforms, such as autonomous vehicles\cite{yang20203dssd, Zheng_2021_CVPR}, robotics\cite{whitty2010autonomous}, and virtual reality (VR)\cite{stets2017visualization}. 3D sensors, such as LiDARs, can generate point clouds at a rapid speed. A typical point cloud consisting of 120,000 points is generated every 1/10th of a second in autonomous driving\cite{velodyne-64}, posing a significant processing workload for embedded systems. 

\begin{figure}[tb]
    \centering
    \includegraphics[width=\linewidth]{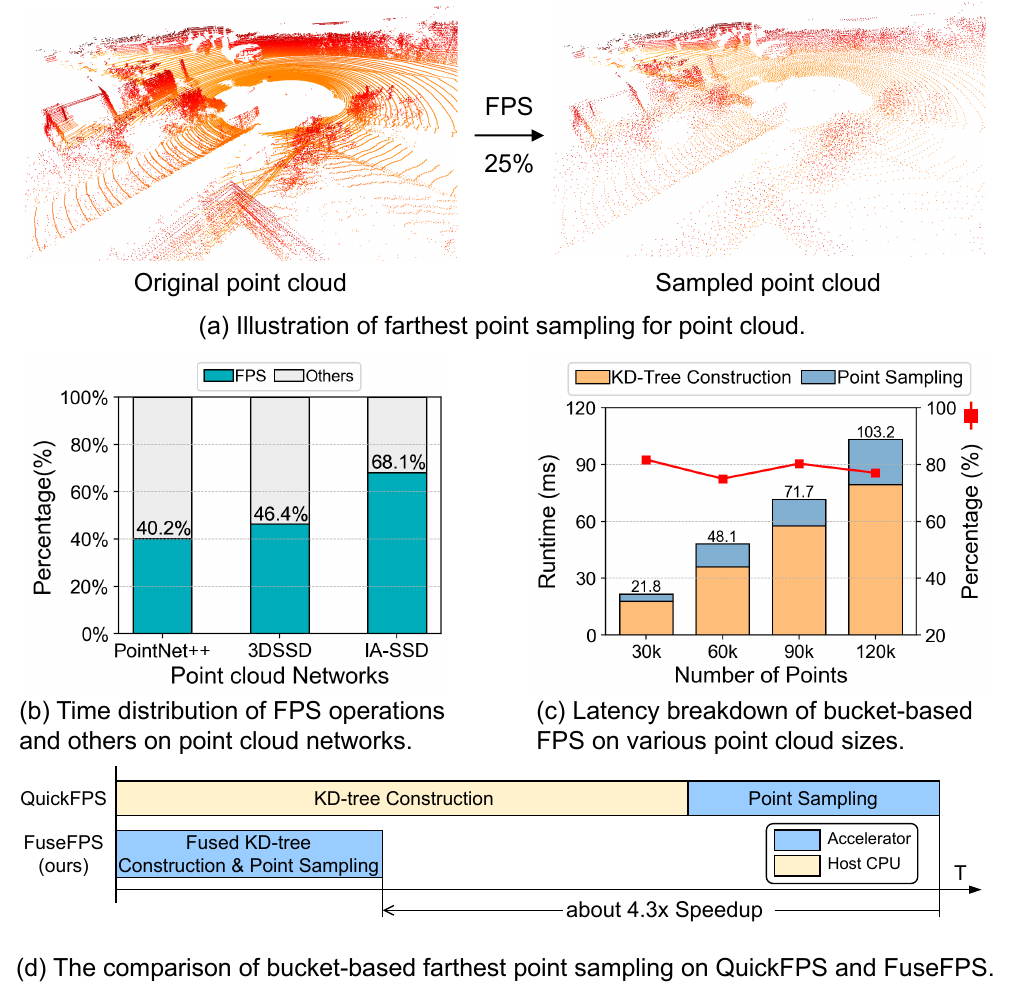}
    \caption{(a) Illustration of FPS for point cloud. (b) Characterize performance of FPS on point cloud neural network. (c) Latency breakdown of BFPS on various point cloud sizes. The red line represents the percentage of the KD-tree construction for total execution time. (d) The comparison between QuickFPS and FuseFPS for execution bucket-based farthest point sampling. }
    \label{fig:fps}
    \vspace{-0.4cm}
\end{figure}

Downsampling is an indispensable step in point cloud processing\cite{dinesh2022point}. Farthest point sampling (FPS)\cite{DBLP:journals/tcs/Gonzalez85}, which can preserve the original spatial characteristics, is the most commonly used algorithm. Fig.~\ref{fig:fps}(a) illustrates the farthest point sampling for the point cloud. Nowadays, FPS kernel is commonly in point cloud neural networks\cite{DBLP:conf/nips/QiYSG17, yang20203dssd, Zheng_2021_CVPR}. However, the execution of FPS incurs heavy memory accesses and high computational costs, becoming a performance bottleneck in point cloud networks. As shown in Fig.~\ref{fig:fps}(b), FPS can account for 40\% to 70\% of the total runtime of point cloud networks on high-end GPU. For a large-scale point cloud ($N\sim10^6$), the conventional FPS takes up to 200 seconds to sample 10\% points on a single GPU\cite{DBLP:conf/cvpr/Hu0XRGWTM20}, demonstrating the importance for accelerating the FPS for real-time point cloud processing.

Prior studies have proposed several domain-specific hardware accelerators to improve the performance of FPS. QuickFPS\cite{han2023quickfps} is the recent work for FPS acceleration. It proposes a bucket-based farthest point sampling (BFPS) accelerator, which employs the KD-tree to reduce unnecessary memory accesses and computation costs. However, QuickFPS requires the aid of the host CPU for KD-tree construction, which leads to a bottleneck in the computing system, specifically in battery-powered embedded and mobile platforms. We profile the performance of BFPS on the Jetson AGX Xavier development board. The KD-tree construction is implemented from popular FLANN library\cite{Flann}. The point cloud size varies from 30,000 to 120,000\footnote{The same sample rate is set 25\% in default, unless specified otherwise.}. Fig.~\ref{fig:fps}(c) shows the latency breakdown of the BFPS. As we can see, the KD-tree construction stage is the predominant contributor to execution time. It takes about 80\% of total runtime on different scales of point clouds, clearly indicating the need for accelerating KD-tree construction.

To tackle such a dilemma, we present FuseFPS, an efficient architecture and algorithm co-design for bucket-based farthest point sampling. The key idea of FuseFPS is to fuse the KD-tree construction stage and point sampling stage efficiently. To achieve that, we first propose a hardware-friendly sampled-driven KD-tree construction algorithm. The algorithm builds the KD-tree step by step during the point sampling, reducing memory accesses further. In addition, The algorithm uses the arithmetic mean value to split the point cloud into buckets, which is more hardware-friendly for the accelerator. Then, we design an efficient accelerator named FuseFPS for bucket-based farthest point sampling. FuseFPS can offload the entire kernel at a low hardware cost.

We implement and synthesize the FuseFPS hardware in RTL and evaluate FuseFPS on three standard workloads. Compared to state-of-the-art accelerator QuickFPS, FuseFPS achieves about 4.3$\times$ and about 6.1$\times$ improvements on speed and power efficiency, respectively.

\section{Background and Related Works}\label{sec:background}

\subsection{Point Cloud Data}

A point cloud is a collection of points that represent 3D scenes or objects. Each point is represented by $<x,y,z>$ in the Cartesian coordinates system. Unlike images, the point cloud directly preserves the 3D geometric information of a scene and the spatial relationship between objects of interest. The point cloud is becoming increasingly prevalent in various emerging AI systems, such as autonomous vehicles\cite{yang20203dssd, Zheng_2021_CVPR}, robotics\cite{whitty2010autonomous}, and virtual reality (VR)\cite{stets2017visualization}. 

\subsection{Bucket-based Farthest Point Sampling}\label{subsec:bfps}

Farthest Point Sampling (FPS)\cite{DBLP:journals/tcs/Gonzalez85} is one of the most commonly used downsampling operations for point cloud analytics. It iteratively generates a subset of the point cloud by selecting the point that is farthest from the previous sampled points in each iteration. As it can effectively preserve the original spatial characteristics\cite{DBLP:journals/tip/EldarLPZ97}, FPS is commonly employed in various point cloud neural networks \cite{DBLP:conf/nips/QiYSG17, yang20203dssd, Zheng_2021_CVPR}. 

FPS always lies in the critical path of the point cloud-based machine perception. However, the heavy memory accesses and high computation costs make FPS a performance bottleneck in computing systems. It can account for 40\% to 70\% of the total runtime of point cloud networks on high-end GPU. When processing a large-scale point cloud ($N\sim10^6$), it takes up to 200 seconds to sample 10\% points\cite{DBLP:conf/cvpr/Hu0XRGWTM20}. 

\begin{figure}[tb]
    \centering
    \includegraphics[width=\linewidth]{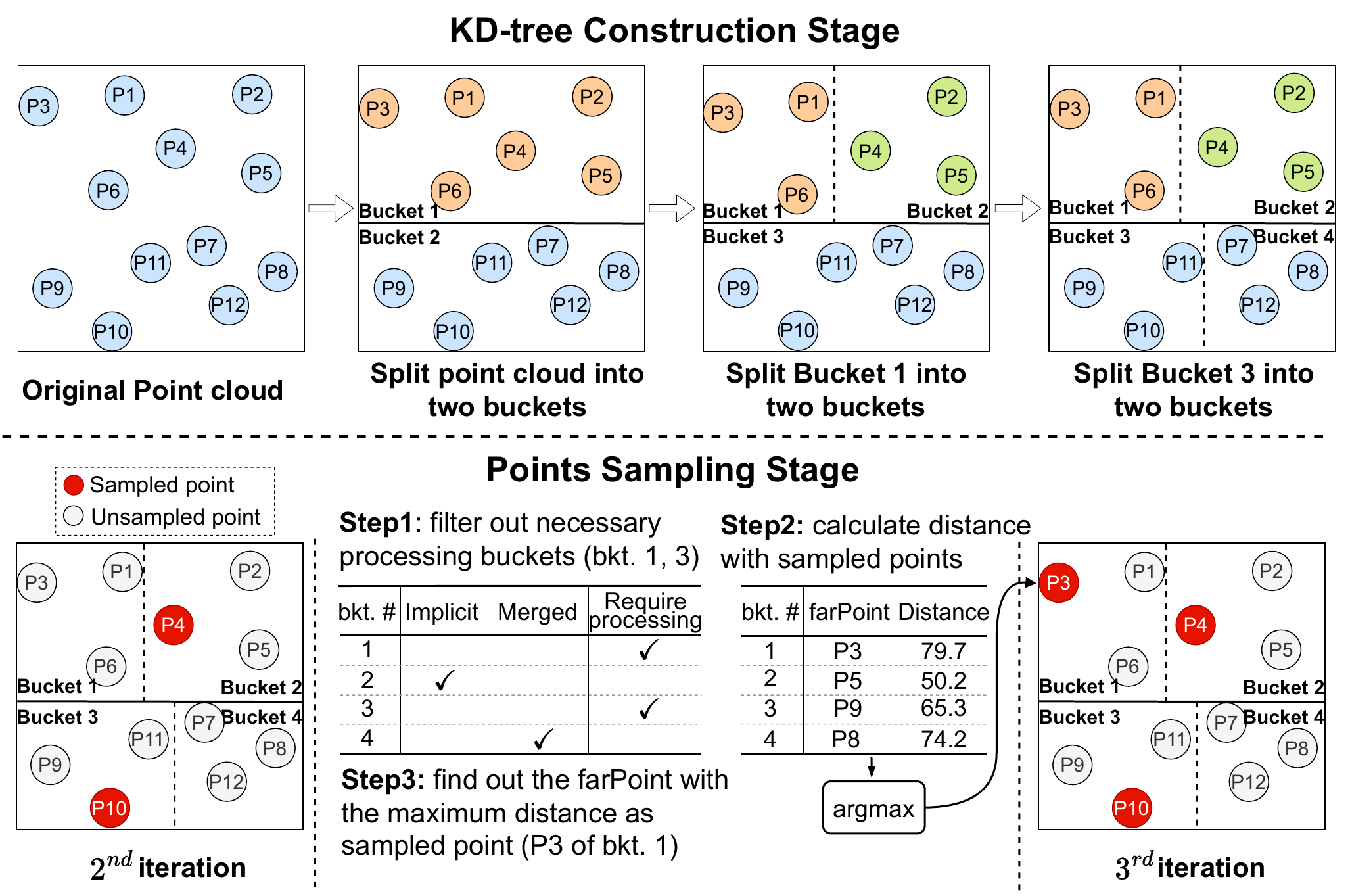}
    \caption{Illustration of the bucket-based farthest point sampling algorithm. The point cloud consists of 12 points. At the point sampling stage, we assume two points (P4, P10) has been sampled.}
    \label{fig:bucket-basedFPS}
    \vspace{-0.3cm}
\end{figure}

Recent work proposes an optimized algorithm for FPS named bucket-based farthest point sampling (BFPS)\cite{han2023quickfps}. BFPS consists of two key stages, KD-tree construction and point sampling. It employs a KD-tree to split the point cloud into multiple smaller buckets. Each bucket contains a part of the point cloud. Then, BFPS designs pruning mechanisms to only process a few necessary buckets during the point sampling stage, significantly reducing unnecessary memory accesses. Fig.~\ref{fig:bucket-basedFPS} gives an example of BFPS. At the KD-tree construction stage, four buckets are created by three-times splitting operations according to the median value of one dimension ($x$ or $y$ dim in this example) step by step. At the point sampling stage, to select the third sampling point, BFPS filters out the necessary buckets (bucket 1, 3) that do not satisfy pruning computation condition among all buckets. Then, it calculates the distance with reference points for each point of each necessary bucket. Finally, the point $P3$ is chosen due to the maximum distance. 

Compared to the conventional FPS, BFPS can generate a sample point by processing only a small portion of the point cloud in each iteration, resulting in significant reduction of memory access. However, the current implementation of BFPS employs a domain-specific accelerator solely for point sampling stage, while relying on the host CPU to split  the point cloud into multiple independent buckets (i.e. the KD-tree construction stage). We break down the runtime on various point cloud sizes. In Fig.~\ref{fig:fps}(c), as the point sampling stage has been well-optimized by the hardware, the KD-tree construction stage takes about 80\% of the total time. This observation highlights the need for further optimizations in the KD-tree construction stage to enhance the overall efficiency and performance of the BFPS algorithm.

\subsection{Related Works}

\textbf{Farthest point sampling accelerator} PointAcc\cite{PointAcc} and QuickFPS\cite{han2023quickfps} are two recent works that support FPS acceleration. PointAcc designs a 6-stage pipeline MPU to execute the conventional FPS algorithm. However, it suffers from heavy memory accesses when the point cloud size exceeds the on-chip memory capacity, which limits it to processing small-scale point clouds. QuickFPS proposes a bucket-based farthest point sampling (BFPS) algorithm and designs a domain-specific accelerator. However, QuickFPS relies on the host CPU to construct the KD-tree, which becomes a performance bottleneck, as discussed in Sec.~\ref{subsec:bfps}.

\textbf{KD-tree construction} Prior works have proposed several optimizations for accelerating KD-tree construction on CPU\cite{cao2020new}\cite{choi2010parallel}, GPU\cite{hu2015massively, wehr2018parallel}. For ASIC, QuickNN\cite{DBLP:conf/hpca/PinkhamZZ20} designs an independent engine. It first selects a subset of point clouds to build a tiny KD-tree. Then, the engine inserts the rest points into the tree node. FastTree\cite{liu2015fasttree} develops a fully parallel construction algorithm and evaluates it as an ASIC implementation. The total area of FastTree is about 1.4 $mm^2$@28nm. Moreover, both QuickNN and FastTree design sorting units to split the point cloud, which often incur high hardware costs. 

In brief, BFPS shows potential for efficiently executing farthest point sampling. However, the current accelerator relies on the host CPU to construct the KD-tree, which becomes a critical bottleneck. Although prior works have designed architectures to accelerate the KD-tree construction. These designs suffer from high hardware costs. Unlike previous works, we design a hardware-friendly mechanism to split the point cloud without sorting. In addition, we efficiently fuse the KD-tree construction with the point sampling stage to reduce memory access further. As a result, FuseFPS gains $4.3\times$ performance improvement with little hardware cost (0.08$mm^2$@28nm).

\section{Fusing the KD-tree Construction with Point Sampling} \label{sec:algorithm}

In this section, we propose an algorithm and architecture co-design mechanism that fuses the KD-tree construction with point sampling to reduce memory access and improve the performance of bucket-based farthest point sampling. To achieve that, we first present a hardware-friendly bucket splitting method (Sec.~\ref{subsec:hardware-friendly}). Then, we introduce the fusion technique named sampled-driven KD-tree construction (Sec.~\ref{subsec:fusion}).

\subsection{Hardware-friendly bucket splitting method} \label{subsec:hardware-friendly}

The KD-tree construction is a recursive process. The key action is splitting a bucket into two sub-buckets. The conventional KD-tree construction method builds a complete binary tree. Each bucket contains equal number of points. Therefore, sort-based KD-tree construction is the most common implementation. However, sorting the large-scale point cloud is time-consuming and often incurs high hardware costs. Especially when the bucket size exceeds the on-chip buffer capacity, the accelerator needs to translate the point cloud data between off-chip memory and on-chip buffer several times.

To address this issue, we employ the arithmetic mean for splitting the point cloud. It only needs to load the bucket once and count the summation of coordinates for each dimension of the points within the bucket, which shows significant hardware efficiency. The implementation of arithmetic mean necessitates a minimal set of hardware components, including accumulators, summation buffers, and a divider.

\begin{figure}[t]
    \centering
    \includegraphics[width=\linewidth]{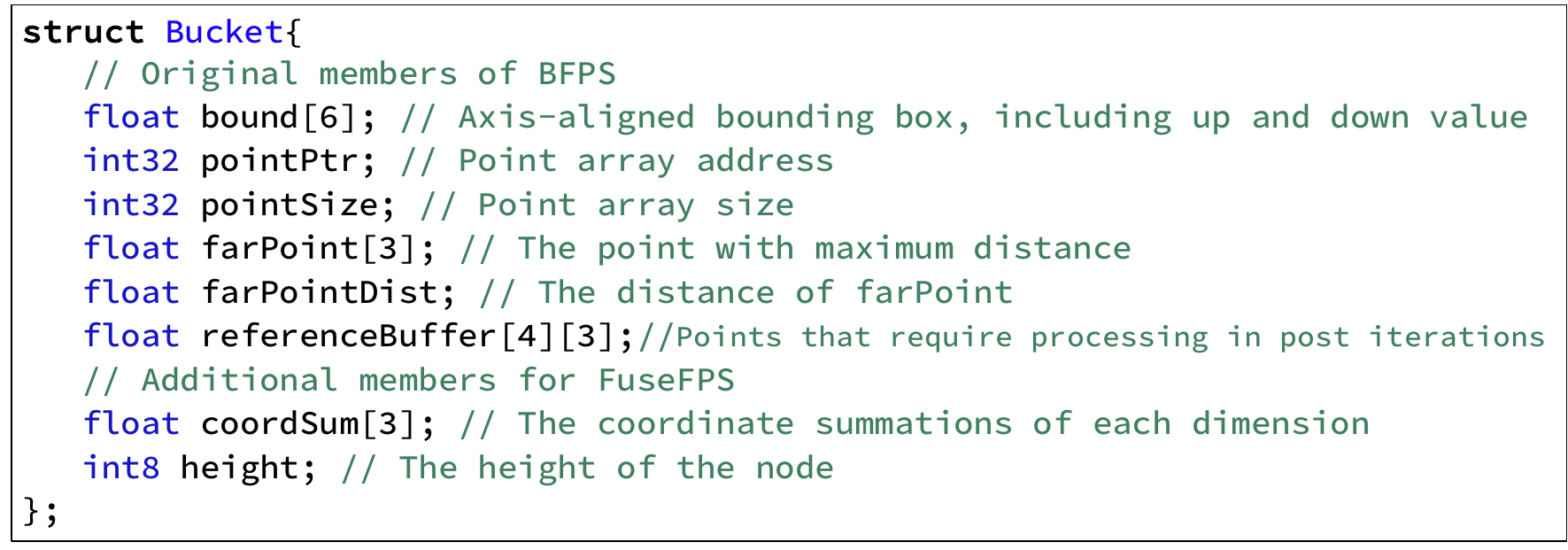}
    \caption{Illustration of the bucket data structure.}
    \label{tab:datastructure}
    \vspace{-0.2cm}
\end{figure}

Furthermore, to support arithmetic mean-based KD-tree construction, we enhance the \texttt{bucket} data structure of BFPS. Fig.~\ref{tab:datastructure} demonstrates the members of the bucket structure, which include two additional members. The $coordSum$ stores the coordinate summations of each dimension for all points within the bucket. The $height$ identifies whether this bucket is required to split further.

\subsection{Sampled-driven KD-tree Construction Algorithm}\label{subsec:fusion}

\begin{figure}[tb]

\removelatexerror

\begin{algorithm}[H]
\begin{small}
\SetKwData{Left}{left}
\SetKwData{This}{this}\SetKwData{Up}{up} 
\SetKwFunction{satifyMerged}{satifyMerged}
\SetKwFunction{satifyImplicit}{satifyImplicit} 
\SetKwFunction{getMinDist}{getMinDist} 
\SetKwFunction{getFarthestPoint}{getFarthestPoint} 
\SetKwFunction{pushback}{pushback} 
\SetKwInOut{Input}{Input}\SetKwInOut{Output}{Output}
\SetKw{continue}{continue}
\caption{Sampling-Driven KD-tree Construction}\label{alg:sampled-driven}
\Input{Bucket $b$, Reference Points $refPoints$\;}
\Output{Updated KD-tree with two new nodes;}
\textbf{new} Buckets $leftChild$, $rightChild$\;
$dim \leftarrow$ \begin{small}$\mathop{\arg\max}\limits_{0\leq i<3}(b.bound[i].up-b.bound[i].down)$\end{small}\;
$splitValue\leftarrow \frac{b.coordSum[dim]}{b.pointSize};~$\tcp{arithmetic mean}
\ForEach {point $p$ in $b$}{
    \tcp{compute distance with $refPoints$}
	$p.dist\leftarrow$ getMinDist($p$, $\mathit{refPoints}$)\;
	\tcp{split $b$ into two child buckets}
	\eIf {$p.coord[dim]$ $<$ $splitValue$} {
		\textbf{updateBucket}($p$, $\mathit{leftChild}$)\;
	}{
		\textbf{updateBucket}($p$, $\mathit{rightChild}$)\;
	}
}
$\mathit{leftChild.height} \leftarrow b.height+1$\;
$\mathit{rightChild.height} \leftarrow b.height+1$\;
\textbf{Delete} Bucket $\mathit{b}$ and \textbf{Append} $\mathit{leftChild, rightChild}$\;

\BlankLine 
\SetKwFunction{UpdateBucket}{\textbf{updateBucket}}
\SetKwProg{Fn}{Function}{:}{}
\Fn{\UpdateBucket{Point $p$, Bucket $b$}}{
    $b.points$.append($p$)\;
    $b.pointSize += 1$\; 
    \For{i in range(0,3)}{
    $\mathit{b.coordSum[i]} += p.coord[i]$\;
    \begin{small}$\mathit{b.bound[i].up = \max(b.bound[i].up, p.coord[i])}$\end{small}\;
    \begin{footnotesize}$\mathit{b.bound[i].down = \min(b.bound[i].down, p.coord[i])}$\end{footnotesize}\;
    }
    \If{$p$.$dist$ $>$ $b$.$farPointDist$}{
        $b$.$farPoint$ $\leftarrow$ $p$\; $b$.$farPointDist$ $\leftarrow$ $p$.$dist$\;
    }
}
\end{small}
\end{algorithm}
\vspace{-0.4cm}
\end{figure}

We propose a sampling-driven KD-tree construction algorithm that efficiently fuses the KD-tree construction with point sampling. In general, We intermittently construct the KD-tree. When a bucket, denoted as $b$, needs to update the distance with reference points and its \texttt{height} is smaller than the given threshold, the algorithm computes the distance for every points within the bucket and split the bucket into two small buckets simultaneously. Otherwise, the algorithm only calculates the distance since the current bucket does not require further splitting. Therefore, the KD-tree remains unchanged. Algorithm~\ref{alg:sampled-driven} shows the pseudo-code of the algorithm. The algorithm first creates two initial sub-buckets, namely $leftChild$ and $rightChild$, which are used to represent the child nodes of the current bucket. As shown in lines 2-3, the dimension for splitting the bucket is determined based on the maximum range of the bounding box across all three dimensions. The arithmetic mean of the corresponding dimension is set to $splitValue$. Then, the bucket points are loaded from off-chip memory. As shown in lines 4-9, for each point within the bucket, the distance between that point and the $refPoints$ is calculated. Subsequently, the point is allocated to one of the two sub-buckets based on its coordinate value along the determined splitting dimension. The lines 14-20 further describes the updating for the sub-bucket when a new point is inserted. After processing all the points within the current bucket $b$, the bucket $b$ is removed, and the two sub-buckets $leftChild$ and $rightChild$ are inserted into the KD-tree.

\begin{figure}[tb]
    \centering
    \includegraphics[width=\linewidth]{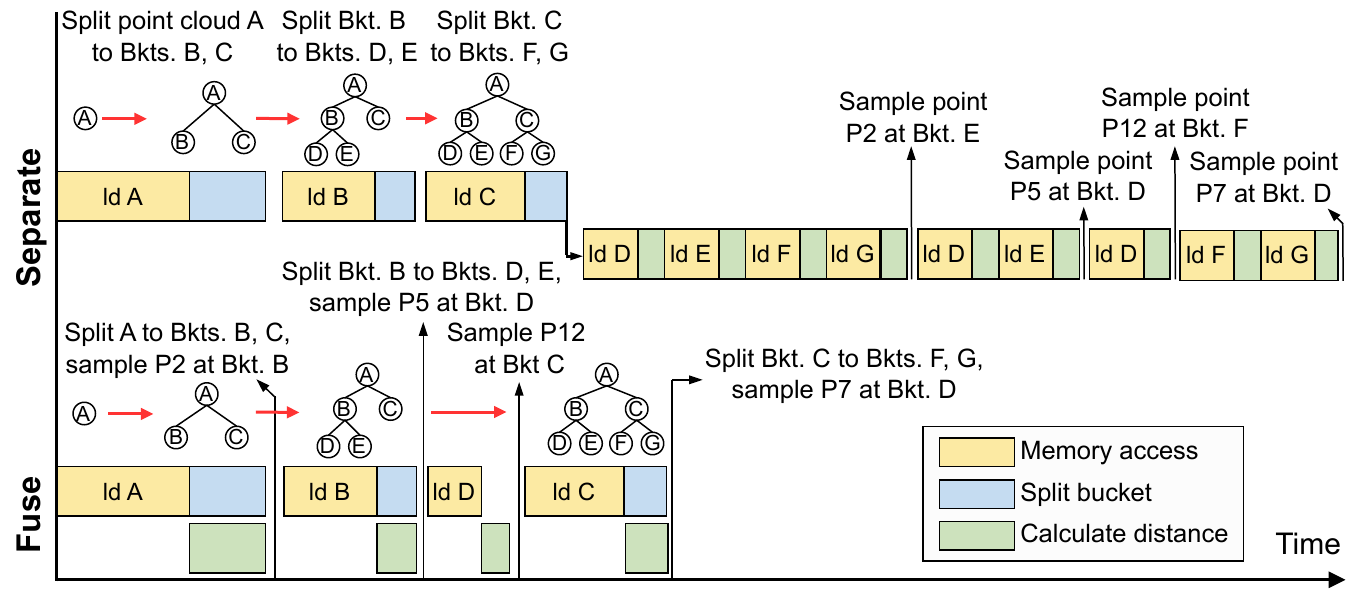}
    \caption{Comparison of separated KD-tree construction with point sampling and fused KD-tree construction with point sampling. The workload is splitting a point cloud (A) into four buckets (D, E, F, G) and sampling four points.}
    \label{fig:compare}
    \vspace{-0.3cm}
\end{figure}

The sampled-driven KD-tree construction algorithm significantly improves performance by reducing off-chip memory access. As mentioned before, the large data size of point clouds and memory-intensive nature of the BFPS result in numerous time-consuming off-chip memory access operations during KD-tree construction and point sampling stages. This affects algorithm efficiency and performance. Our algorithm addresses it by efficiently using the accessed data. In Fig.~\ref{fig:compare}, we use an example to show the benefits of fusing compared to that without fusion. In the fusing solution, the KD-tree construction goes along with point sampling. Compared to the separated solution, the fused solution loads the bucket data once, while performing both the splitting bucket and calculating distance simultaneously. We have conducted an experiment and found that compared to the separated solution, the fusing solution achieves about 16.9\% DRAM access reduction. This is further analyzed in Section~\ref{subsec:sensitivity}.

\section{FuseFPS Accelerator Design}\label{sec:accelerator}

\subsection{Accelerator Overview}
\begin{figure}[tb]
    \centering
    \includegraphics[width=\linewidth]{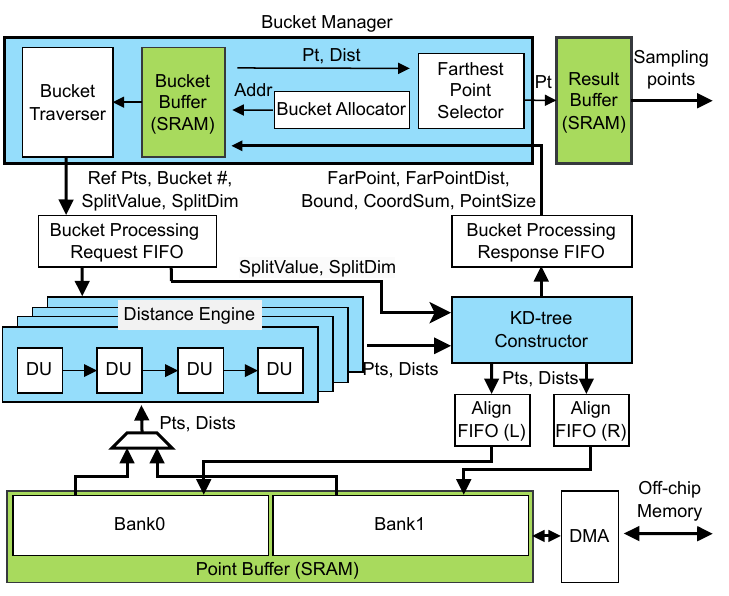}
    \caption{Overview of FuseFPS architecture. }
    \label{fig:fuseFPSarchitecture}
    \vspace{-0.3cm}
\end{figure}

Fig.~\ref{fig:fuseFPSarchitecture} shows the accelerator architecture of FuseFPS, which consists of four primary parts: bucket manager, distance engine, KD-tree constructor and point buffer.  

During the runtime, the bucket manager plays the role of controller of the accelerator. It uses multiple FSM-based modules to issue the bucket processing requests and generate the farthest points. The distance engine consists of several 1D systolic arrays of distance units (DUs). Each distance unit performs the function like $f(p,q) = \min((p-q)^2, p.dist)$, where $p$ is the bucket point from point buffer, $q$ is the reference point from request FIFO. In fact, the point buffer reads 4 points per cycle, and 4 DU-arrays individually process 1 point each. The KD-tree constructor receives the points from the distance engine and issues them to one of two align FIFOs based on each point's coordinate value along the splitting dimension. KD-tree constructor updates the status of the child bucket. When finishing bucket processing, the responses are sent to the bucket manager. After processing all necessary buckets, the farthest point selector of the bucket manager finds out the farpoint with maximum distance as the sampling point and writes it to the result buffer. 

\subsection{Bucket Splitting in FuseFPS}

\begin{figure}[tb]
    \centering
    \includegraphics[width=\linewidth]{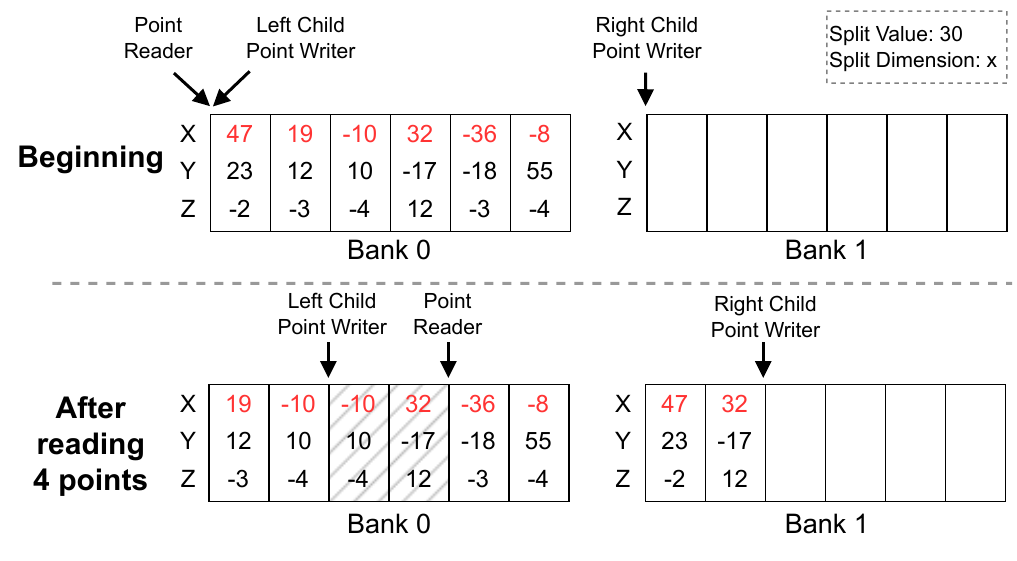}
    \caption{Example of bucket splitting. The bucket is split by the value 30 at x dimension. For simplicity, point buffer only contains 1 point per SRAM row. The SRAM rows marked with gray slashes represent the data within the rows have been read but have not be written with new data.}
    \label{fig:splitbuckets}
    \vspace{-0.3cm}
\end{figure}

Fig.~\ref{fig:splitbuckets} gives an example of bucket splitting. The point buffer consists of two SRAM banks. Both have one read port and one write port. When splitting a bucket into two child buckets, the parent bucket point data is loaded into point bank 0 while keeping bank 1 empty. The read pointer is initialized to the head of bank 0. Two child write pointers are initialized to the heads of bank 0 and bank 1 respectively. 

During the processing, the distance engine reads the point data through the read pointer, computes distance with reference points, and sends it to the KD-tree constructor. The constructor issues the point based on the value at the x dimension. In this example, if the point's value of x dimension is smaller than 30, the point will be issued to align FIFO (L). Then, the aligned points within FIFO are written to SRAM bank. Fig.~\ref{fig:splitbuckets} shows the bank context after reading 4 points.

\section{EVALUATION}

\subsection{Evaluation Setup}\label{subsec:setup}

\textbf{Hardware Implementation} FuseFPS is implemented with Verilog and verified through RTL simulations. We synthesize FuseFPS with Synopsys tools in a 28nm process technology. The clock is set to 1 GHz. The memory compiler generates the SRAMs. We integrate the RTL model with DRAMsim3\cite{DRAMsim3} to model the DRAM4-2400 behaviors and measure the DRAM energy consumption. We generate the Switching Activity Interchange Format (SAIF) files during the simulation and send them to Design Compiler to estimate the power consumption. 

\textbf{Benchmark} Table~\ref{tab:benchmark} lists the three point cloud datasets that are used in the evaluation. The datasets contain various sizes ranging from \num{4e3} to \num{1.2e5}. The scene of the point cloud covers the indoors and spacious outdoors. Following the prior studies\cite{DBLP:conf/nips/QiYSG17, yang20203dssd, Zheng_2021_CVPR}, the sampled rate is set 25\%.

\textbf{Baseline} We employ Jetson AGX Xavier GPU, PointAcc, and QuickFPS accelerator as baselines. In detail, We implement a GPU version of bucket-based farthest point sampling for Jetson AGX Xavier. Our implementation achieves 3.9$\times$ speedup over the state-of-the-art implementation OpenPCDet library\cite{openpcdet2020}. QuickFPS employs Jetson AGX Xavier's CPU to perform the KD-tree construction stage. 

\begin{table}[tb]
\centering
\caption{Evaluation Benchmarks.}\label{tab:benchmark}
\begin{minipage}{\linewidth}
\def\arraystretch{1.5}\tabcolsep 2pt
\begin{tabular}{c|@{~~~}c@{~~~}c@{~~~}c@{~~~}c}
\hline
\textbf{Workload} & \textbf{Point size} & \textbf{Sample rate} & \textbf{Dataset} & \textbf{Scene}\\ \hline
 Small     & \num{4.0e3} & 25\%& S3DIS\cite{S3DIS} & Indoors \\ \hline
 Medium & \num{1.6e4} & 25\%& KITTI\cite{Geiger2012CVPR}  & Outdoors\\ \hline
Large & \num{1.2e5} & 25\%& SemanticKITTI\cite{DBLP:conf/iccv/BehleyGMQBSG19}  & Outdoors\\ \hline
\end{tabular}
\end{minipage}
\end{table}

\subsection{Speedup and Power Efficiency}

\begin{figure}[t]
    \centering
    \includegraphics[width=\linewidth]{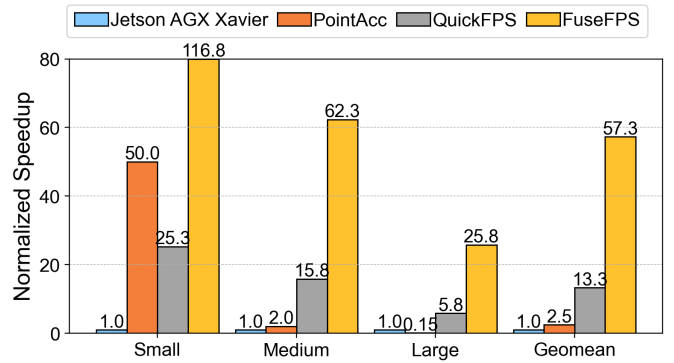}
    \caption{Normalized speedup of FuseFPS with different scales of point clouds. }
    \label{fig:speedup}
    \vspace{-0.3cm}
\end{figure}

\begin{figure}[t]
    \centering
    \includegraphics[width=\linewidth]{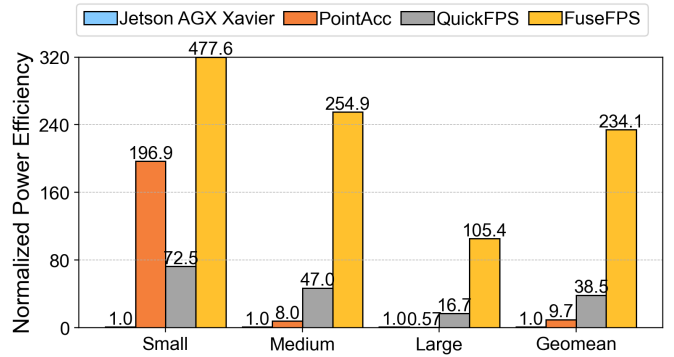}
    \caption{Normalized power efficiency of FuseFPS with different scales of point clouds. }
    \label{fig:energysaving}
    \vspace{-0.3cm}
\end{figure}

We evaluate the performance of FuseFPS on three workloads with different heights of KD-tree. In detail, we set the height of KD-tree for FuseFPS to 6, 7 and 9 for small, medium and large workloads, respectively. For a fair comparison, we adapt the height of KD-tree for Jetson AGX Xavier GPU and QuickFPS to achieve optimal performance.

Fig.~\ref{fig:speedup} presents the speedup of FuseFPS with different scales of point clouds. On average, FuseFPS achieves 57.3x and 23.3x speedup against Jetson AGX Xavier GPU, PointAcc accelerator, respectively. Compared to QuickFPS, which is the most competitive accelerator among those evaluated, FuseFPS obtains 4.3x speedup on average. This is because FuseFPS can effectively perform the KD-tree construction without the aid of the host CPU. PointAcc shows a competitive acceleration effect on the small workload. However, as PointAcc only supports conventional FPS, the large memory footprint makes it even slower than GPU (implement BFPS) when the point cloud size increases.

As illustrated in Fig.~\ref{fig:energysaving}, FuseFPS achieves 234.1x, 24.2x, and 6.1x energy efficiency improvement on average compared to GPU, PointAcc and QuickFPS, respectively. Because of the hardware-friendly KD-tree construction, FuseFPS takes about the overhead of 8\% on-chip power increases than QuickFPS. As shown in Table~\ref{tab:resource}, we present more comprehensive comparisons with QuickFPS.

\subsection{Area and Energy Breakdown}

\begin{figure}[t]
    \centering
    \includegraphics[width=0.9\linewidth]{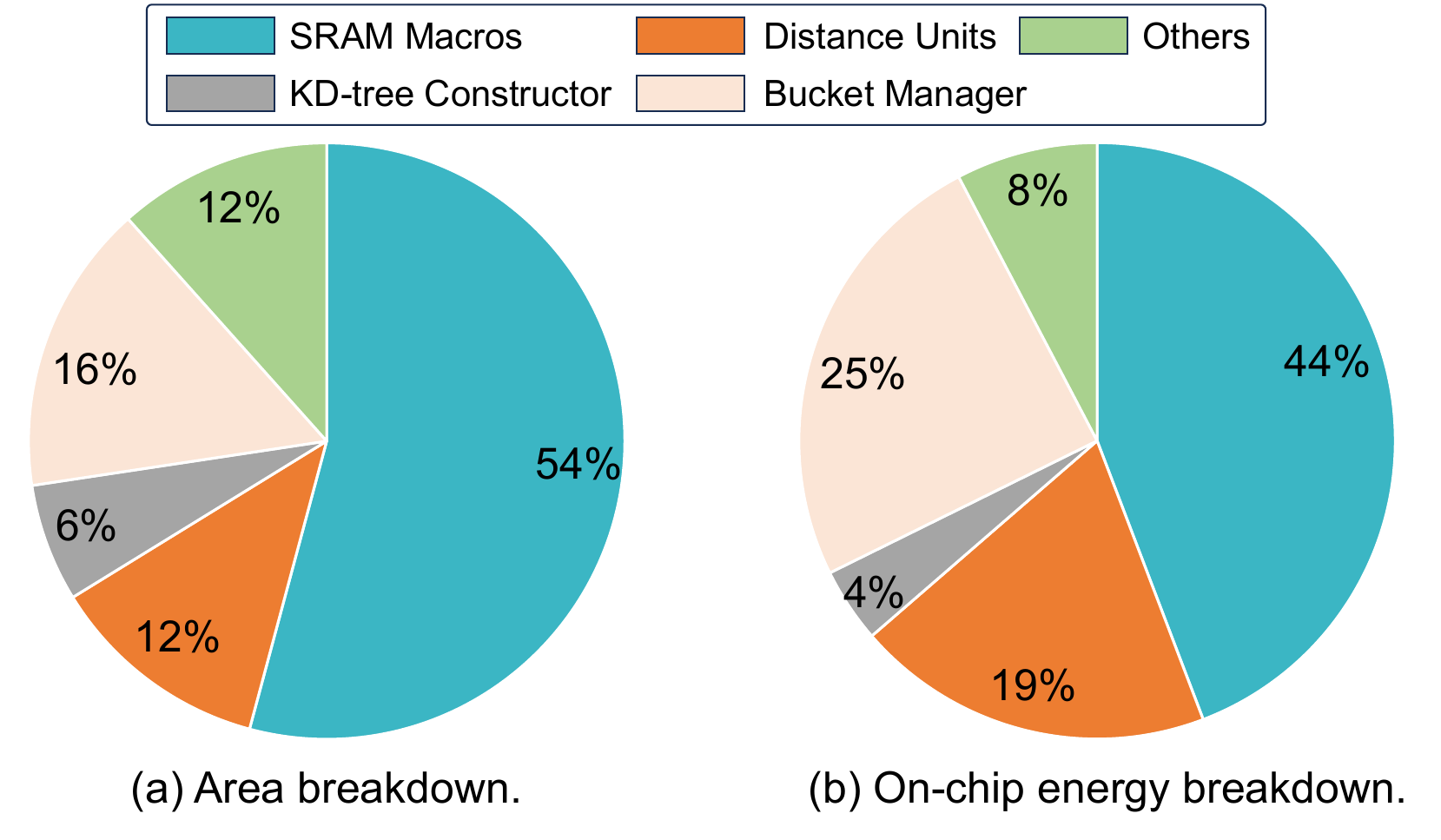}
    \caption{FuseFPS area/energy breakdown. }
    \label{fig:breakdown}
    \vspace{-0.3cm}
\end{figure}

Fig.~\ref{fig:breakdown} presents the area and energy breakdowns of FuseFPS for a large workload, excluding the off-chip memory. As we can see, the overhead of KD-tree constructor is remarkably low, accounting for approximately 6\% of the total area and 4\% of on-chip energy. This exemplifies the lightweight and efficient nature of the hardware-friendly bucket splitting method employed in the KD-tree construction process.

The SRAM buffer and bucket manager are identified as the main contributors to the area utilization. Specifically, FuseFPS comprises a total on-chip SRAM size of 75 KB, with 58.25 KB allocated to the bucket buffer and 16 KB to the point cloud buffer. This configuration allows FuseFPS to support up to 512 bucket instances and 1024 points.

In terms of energy breakdown, the distance engine and SRAMs contribute to 19\% and 44\% of the on-chip components' energy consumption, respectively. According to the DRAM simulator, the off-chip memory consumes approximately 11 mJ of energy during the execution.  In all, the energy consumption of FuseFPS is well-suited for embedded systems, where energy efficiency is a critical consideration.

\begin{figure}[t]
    \centering
    \includegraphics[width=\linewidth]{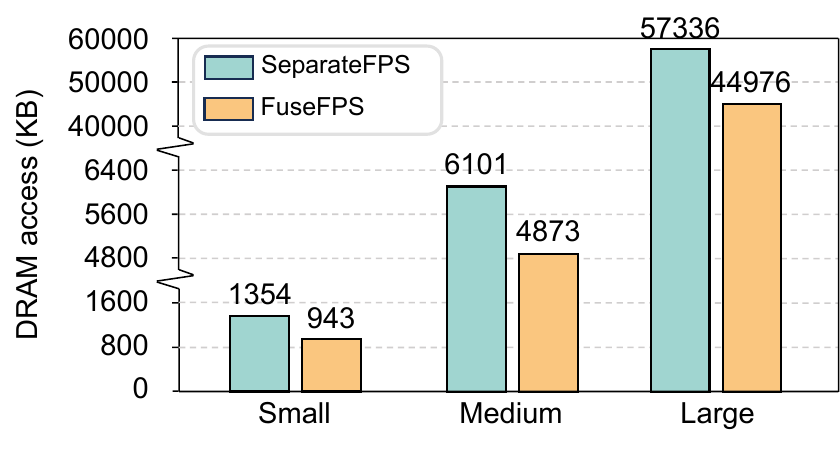}
    \caption{External memory access of three workload on SeparateFPS and FuseFPS. The SeparateFPS represents a hardware accelerator which separately executes the KD-tree construction and point sampling. }
    \label{fig:memoryaccess}
    \vspace{-0.3cm}
\end{figure}

\begin{table}[tb]

\caption{Comparison of Area, On-chip Power, Speedup, Power Efficiency (P.E) and Area Efficiency (A.E) improvement Among FuseFPS and QuickFPS.}\label{tab:resource}
\def\arraystretch{1.5}\tabcolsep 2pt
\def\thefootnote{a}\footnotesize
\begin{tabular}{c | c c c c c c c}
\hline
\textbf{} & \textbf{\makecell[c]{Area\\($mm^2$)}} &  \textbf{\makecell[c]{Power\\(mW)}} & \textbf{\makecell[c]{SRAM\\(KB)}}  & \textbf{\makecell[c]{Frequency\\(GHz)}} &  \textbf{Speedup} & \textbf{P.E} & \textbf{A.E} \\ \hline
\textbf{QuickFPS} & 0.596 &  142  &  66 & 1.0 & $1\times$ & $1\times$ & $1\times$ \\ \hline
\textbf{FuseFPS} & 0.672  &  154  &  75 & 1.0 & $4.3\times$ & $6.1\times$ & $3.8\times$\\ \hline
\end{tabular}
    \vspace{-0.3cm}
\end{table}

\subsection{Sensitivity on Fusion}\label{subsec:sensitivity}

We have conducted additional experiments to gain deeper insights into the memory access reduction resulting from fusion. A hardware accelerator named SeparateFPS is implemented, which separately executes the KD-tree construction and point sampling (like the separate solution in Fig.~\ref{fig:compare}). The height of KD-tree are set to 6, 7 and 9 for both accelerators in small, medium and large workloads. We first count the number of sampled points by FuseFPS upon completing the KD-tree construction. Then we set SeparateFPS to sample an equal number of points. The DRAM access is account for comparisons. As shown in Fig.~\ref{fig:memoryaccess}, Compared to SeparateFPS, FuseFPS achieves 16.9\% DRAM access reduction on average, which further improves the performance of FuseFPS.

\section{Conclusion} \label{sec:conclusion}

This paper presents FuseFPS, a new algorithm and architecture co-design for the farthest point sampling. We fuse the KD-tree construction stage into the point sampling stage. In this way, FuseFPS can offload the entire kernels on the accelerator. Extensive evaluation experiments show that FuseFPS delivers significant speedup and power efficiency over GPU and state-of-the-art point cloud accelerator QuickFPS.

\bibliographystyle{IEEEtran}
\bibliography{references}

\end{document}